%% file: main.tex
\documentclass[a4paper,twoside]{article}

\usepackage{epsfig}
\usepackage{subcaption}
\usepackage{calc}
\usepackage{amssymb}
\usepackage{amstext}
\usepackage{amsmath}
\usepackage{amsthm}
\usepackage{multicol}
\usepackage{pslatex}
\usepackage{apalike}
\usepackage{algorithm2e}
\usepackage[bottom]{footmisc}
\usepackage{hyperref}
\usepackage{multirow} 
\usepackage{booktabs}
\usepackage{comment}
\usepackage{float}
\usepackage{xurl}   

\usepackage{SCITEPRESS}     

\usepackage{dblfloatfix} 
\begin{document}



\title{To Be or Not to Be (in the EU): Measurement of Discrepancies Presented in Cookie Paywalls}
 \author{\authorname{Andreas Stenwreth$^*$\sup{1}\orcidAuthor{0009-0000-4466-8561}, Simon Täng$^*$\sup{1}\orcidAuthor{0009-0007-4340-4370} and Victor Morel\sup{1}\orcidAuthor{0000-0001-9482-8906}}
 \affiliation{\sup{1}Department of Computer Science and Engineering, Chalmers University of Technology, Gothenburg, Sweden}
 \email{\{andstenw, simonta, morelv\}@chalmers.se}
 }

\keywords{Cookie Banners, Cookie Paywalls, Pay-or-okay, GDPR, Web Measurements.}

\abstract{Cookie paywalls allow visitors to access the content of a website only after making a choice between paying a fee (paying option) or accepting tracking (cookie option). The practice has been studied in previous research in regard to its prevalence and legal standing, but the effects of the clients' device and geographic location remain unexplored. To address these questions, this study
explores the effects of three factors: 1) the clients' browser, 2) the device type (desktop or mobile), and 3) the geographic location on the presence and behavior of cookie paywalls and the handling of users' data. 
Using an automatic crawler on our dataset composed of 804 websites that present a cookie paywall, we observed that the presence of a cookie paywall was most affected by
the geographic location of the user. We further showed that both the behavior of a cookie paywall and the processing of user data are impacted by all three factors, but no patterns of significance could be found. Finally, an additional type of paywall was discovered to be used on approximately 11\% of the studied websites, coined the ``double paywall'', which consists of a cookie paywall complemented by another paywall once tracking is accepted.}

\onecolumn \maketitle \normalsize \setcounter{footnote}{0} \vfill
\def\thefootnote{*}\footnotetext{Andreas Stenwreth and Simon Täng are co-first authors with equal contribution and imporantce.}\def\thefootnote{\arabic{footnote}}
\input{include/Introduction}

\input{include/Background}

\input{include/Method}


\input{include/Results}

\input{include/Discussion}

\input{include/Conclusion}

\vspace*{-.5cm}
\bibliographystyle{apalike}
{\small
\bibliography{Paper/references}}


\end{document}

%% file: include/Introduction.tex
\section{\uppercase{Introduction}}
\label{sec:introduction}

\begin{figure}[!ht]
    \centering
    \includegraphics[width=\linewidth]{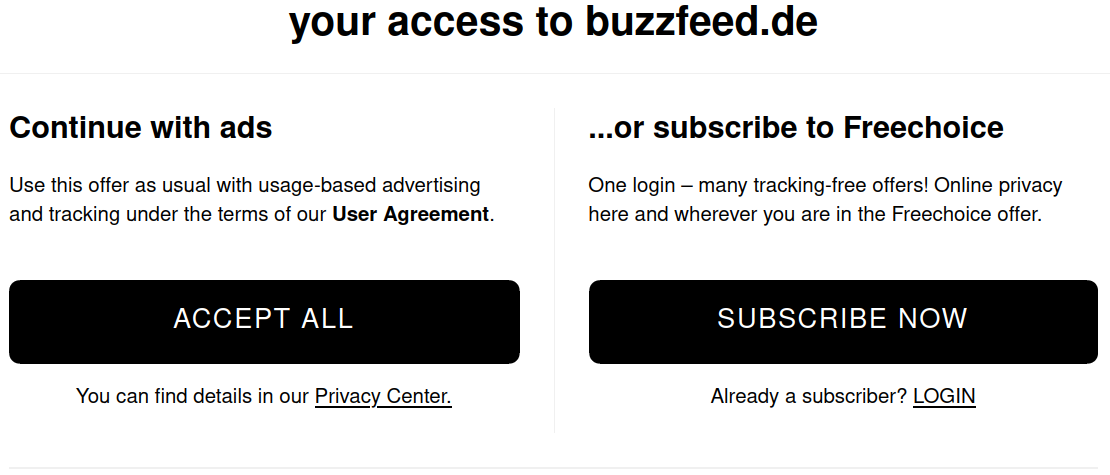}
    \caption{A cookie paywall with a cookie option and a pay option on \url{https://www.buzzfeed.de/}.}
    \label{fig:cookie-paywall}
\end{figure}

Cookie paywalls, sometimes denoted ``pay-or-tracking wall'', ``accept-or-pay cookie banner'' or ``pay-or-okay banner'', allow visitors to access the content of a website only after making a choice between paying a fee or accepting tracking~\cite{morel_your_2022}. This practice, illustrated in Figure~\ref{fig:cookie-paywall}, 
has recently triggered interest from regulators, which has resulted in Meta being charged for breaching the Digital Markets Act~\cite{meta_DMA_Charged} after introducing a pay-or-okay model on Facebook and Instagram.
As the pay-or-okay model is put under review, data is needed to give a thorough basis for regulators and legislative decision-making. 
However, the technical aspects of this practice remain understudied. Notably, previous studies have been limited in scope in regard to the plurality of browsers and environments (e.g. desktop or mobile)~\cite{morel_your_2022,morel_legitimate_2023,rasaii_thou_2023,mueller-tribbensee_paying_2024}. 
Nevertheless, the browser used has been observed to affect
the presence of a paywall on a website~\cite{morel_your_2022}, prompting the need for further studies on the correlation between factors such as browser, environment, geographic location, and a website's behavior. 

Thus, this paper addresses the following research questions:
(1) How is the prevalence of cookie paywalls affected by the browser, device type and geographic location of the user? 
(2) Do the browser, device type and geographic location affect how much of a website can be accessed before a cookie paywall appears?
(3) Do the browser, device type and geographic location affect how and by whom user data is processed?

During this study, we discovered a new type of paywall. 
This ``double paywall'' initially presents a cookie paywall, which is later complemented by an additional paywall after tracking (often performed via the use of cookies) has been accepted. The second paywall is usually only presented on certain parts of a website, such as a ``subscription only'' article, thus restricting access to these parts. 
We contacted a legal scholar, who deems this practice to be of interest from a data protection and consumer protection point of view. We therefore argue that quantitative insights regarding the adoption of this practice will provide a valuable basis for further legal examination. 
As a result, this paper also studies the research question:
(4) How widespread is the occurrence of double paywalls? 

To answer these questions, we develop a crawler to extract data from 804 websites using 26 different combinations of operating system, web browser and geographic location. We manually confirm that these 804 websites use a cookie paywall. The main contributions of this paper can be summarized as follows:
\begin{itemize}
    \item \textbf{The largest cookie paywall dataset to date:} We combine the datasets from previous studies and other sources to create a dataset of 804 websites with confirmed cookie paywalls. This is, to the best of our knowledge, the largest dataset of cookie paywalls to date, doubling the number of cookie paywalls found by previous studies.
    \item \textbf{The choice of user client presents discrepancies in cookie paywalls:} We measure cookie paywalls in terms of numbers, third party vendors, legal purposes for data processing and what user interaction is needed before they are displayed, and search for discrepancies between combinations of operating system, browser and geographic location. We found that the geographic location of the user has a significant impact on whether a cookie paywall is presented, and that some Consent Management Platforms (CMPs) are more frequent than others among the websites that show discrepancies. Both the choice of operating system and browser also correlate with the presence of cookie paywalls, and all three factors seem to impact how users' data is processed.
    \item \textbf{The first dataset of double paywalls:} We introduce the first dataset of double paywalls (93 websites). This data has already been presented during a panel discussion organized by the European Data Protection Supervisor in July 2024. 
\end{itemize}

%% file: include/Background.tex
\section{\uppercase{Background and Related Work}}

Here we describe the relevant legal notions regarding collection and processing of personal data in the European Union (EU).
We further outline an industry standard for enforcing legal requirements of the applicable laws and regulations, as well as actors integral to the technical implementation of cookie paywalls.

\subsection{Legal Landscape}
\label{gdpr_epd}
In the EU, two main data protection laws, namely the General Data Privacy Regulation (GDPR)~\cite{GDPR2016a} and the ePrivacy Directive (ePD)~\cite{parliament_directive_2002}, regulate the processing, storing, and management of personal data (for the GDPR), and of cookies more precisely (for the ePD).
In this context, personal data is ``any information relating to an identified or identifiable natural person'', and because a significant amount of data on the web is considered personal data (typically, data stored in web cookies), organizations collecting such data (i.e. controllers) must abide by the requirements laid out in these texts.
They notably require that controllers choose a \textbf{legal basis} (i.e. a legally valid high-level reason), such as the \textit{consent} of the user or the \textit{legitimate interest} of the organization~\cite[Recital 36]{GDPR2016a}, and a specific purpose~\cite[Parag. 121]{edpb_guidelines_2020} for the processing of personal data. 
When using cookies for the intention of tracking or targeted advertising, the GDPR and ePD state that user \textit{consent is the only applicable legal basis}~\cite[Art. 5(3)]{parliament_directive_2002} and users must be notified of any tracking and be able to act, opt out, or leave before any tracking is initiated~\cite[Art. 6]{GDPR2016a}. 
The GDPR also stipulates five requirements for consent to be valid, it must be: (1) given by clear affirmative act, (2) freely given, (3) specific, (4) informed, and (5) give an unambiguous indication of the user’s agreement~\cite[Art. 4(11)]{GDPR2016a}. 

These legal notions form the substrate of the reason why websites present cookie banners: website owners must inform their visitors and elicit their consent in order to lawfully collect their data, at least in the EU.
However, what constitutes a valid consent is still subject to some interpretations, and in the case at hand, poising consent against a fee (i.e. cookie paywalls) has not been ruled illegal although national Data Protections Agencies (DPAs) have divergent opinions on the matter~\cite{morel_legitimate_2023}.

\subsection{Transparency and Consent Framework}
\label{tcf}

The Transparency and Consent Framework (TCF), created by the Interactive Advertising Bureau (IAB), is an industry standard and tool designed to create a standardized experience when making privacy choices on websites~\cite{iab_europe_transparency_2023}. The standard specifies eleven purposes for data processing that can rely on one or both legal bases of consent and legitimate interest (depending on whether the purpose relates to targeted advertising or not). The purposes are: (1) Store and/or access information on a device, (2) Select basic  ads, (3) Create a personalized ads profile, (4) Select personalized  ads, (5) Create a personalized content profile, (6) Select personalized  content, (7) Measure ad performance, (8) Measure content performance, (9) Apply market research to generate audience insights, (10) Develop and improve products, and (11) Use limited data to select content. 
The elicitation of consent and the communication of the purposes and legal bases for data processing is implemented through consent notices, such as cookie banners, commonly provided by Consent Management Platforms (CMPs).

CMPs provide functions including presenting a consent banner to ask for user consent and logging the provided response of a user~\cite{iab_europe_cmp_2024}. After consent is given,\footnote{Note that for legitimate interest, the controller does not have to elicit an affirmative action, only informing users is required.} the CMP regulates the activation of cookies and other technologies based on the given consent in line with the EU data protection regulations. Additionally, CMPs package user preferences in a standardized payload called the Transparency and Consent String (TC String). 

A TC String is an encoded HTTP-transferable string, starting with the character C, that enables communication of transparency and consent information~\cite{iab_tech_lab_transparency_2023}. 
The TC String contains information such as the number of vendors to which data is conveyed based on user consent or legitimate interest, as well as the purposes for which consent and legitimate interest are used as a basis for data processing. The information is passed to all relevant parties including the data subject, the publisher, and the vendors. A TC String is stored in the user's web browser as either a persistent cookie or as an entry in the browser's local storage.

Together with presenting a consent banner, many CMPs also provide integration with cross-site subscription-based models -- provided by companies coined Subscription Management Platforms (SMPs) --, as a way of offering additional revenue streams for websites. 
SMPs such as Content Pass GmbH~\cite{content_pass_gmbh_contentpass_2024} and Traffective GmbH~\cite{traffective_gmbh_freechoice_2024} offer such subscription-based models in the form of contenpass and Freechoice, respectively.
These products provide the option of paying a monthly fee to gain access to all partnered websites without personalized advertisement~\cite{pfau_pur_2023}. 

\subsection{Related Work}
\label{related work}
\cite{rasaii_exploring_2023} developed an automated detection method of cookie banners using various vantage points and both a desktop and a simulated mobile device. The authors examined the effect of the vantage point and device type on the presence of cookie banners and the number of tracking cookies employed. Similarly, \cite{van_eijk_impact_2021} used a VPN to study the impact of user location on cookie notices using an automated detection method.

The first study to investigate the privacy impacts of paywalls was conducted by \cite{papadopoulos_keeping_2020} in which paywalls were categorized based on the restrictions put on the user.
However, \textit{cookie} paywalls were first studied by \cite{morel_your_2022} who manually identified and found cookie paywalls on 13 out of the 2800 websites they studied. The detection process was later automated by \cite{morel_legitimate_2023} and \cite{rasaii_thou_2023} by employing a web crawler using the Mozilla Firefox web browser. Neither of the studies considered the use of different browsers nor examined the role of mobile devices. \cite{morel_legitimate_2023} found 431 websites using cookie paywalls, all of which used the TCF, and \cite{rasaii_thou_2023} found 280 websites. Both studies used language-based approaches to identify whether a website uses a cookie paywall, leveraging paywall-related keywords.

%% file: include/Method.tex
\section{\uppercase{Methodology}}

\begin{figure*}[!ht]
    \small
    \includegraphics[width=\linewidth]{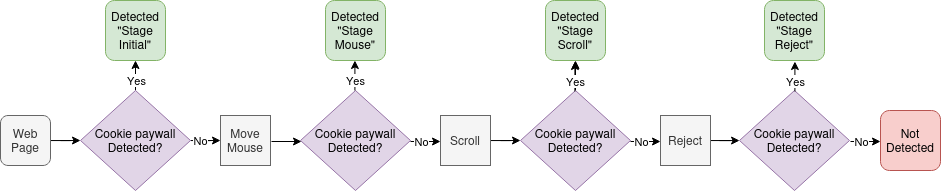}
    \caption{The program flow for detecting cookie paywalls in different stages.}
    \label{fig:flowchart_web_crawler}
\end{figure*}

We present here how we designed the web crawler used to collect data from 804 websites in June 2024. The crawler is built using Selenium~\cite{software_freedom_conservancy_selenium_2024_new} and Appium~\cite{openjs_appium_2024} to allow the use of desktop and mobile browsers to search websites for cookie and double paywalls. It leverages natural text processing and HTML IDs and class names commonly used by cookie paywalls to detect cookie and double paywalls, and additionally collect and analyze TC Strings. Moreover, the crawler is configurable using three factors, the geographic location, operating system and browser, to allow for the comparison of website behavior between different device setups.

\subsection{Factors and Dataset}
\label{sec:dataset}
We use two geographic locations, Gothenburg in Sweden, covered by the GDPR and ePD, and New York in the USA, not covered by European privacy laws. The Swedish location was accessed directly and the USA location was set up using a VPN service. 
Moreover, we use five operating systems: Windows, macOS, Linux, Android and iOS, and the top four most used web browsers: Google Chrome, Safari, Microsoft Edge and Mozilla Firefox~\cite{statcounter_browser_2023}. These combine into 26 different configurations of the crawler. 


The dataset consists of the 431 websites found to be using cookie paywalls by \cite{morel_legitimate_2023}, the 280 websites found by \cite{rasaii_thou_2023}, plus an additional 441 and 215 websites listed on the websites \textit{contentpass.net}~\cite{contentpass_link} and \textit{freechoice.club}~\cite{freechoice_link}, the only SMPs known to the authors during the time of the study. After cleaning the dataset of duplicates and websites that did not present cookie paywalls, these combine into a final dataset of 804 websites\footnote{\url{https://github.com/Trooja/cookie-paywall-discrepancies/blob/main/cookie-paywalls.csv}}.

\begin{table*}[!b]
    \small
    \centering
    \setcounter{table}{1}
    \caption{Examples of syntagm combinations used for detecting cookie paywalls.}
    \begin{tabular}{@{}c|ll@{}}
    \specialrule{.15em}{.075em}{.01em} 
    Language                 & \multicolumn{1}{c}{Pay option} & \multicolumn{1}{c}{Cookie option} \\ \specialrule{.08em}{.0em}{.0em} 
    \multirow{2}{*}{English} & ``subscribe and decline''      & ``reject and subscribe''           \\
                             & ``read ad-free''               & ``with advertising and tracking''  \\ \hline
    \multirow{2}{*}{German}  & ``jetzt abonnieren''           & ``akzeptieren und weiter''         \\
                             & ``bereits abonnent''           & ``weiter mit werbung''             \\ \hline
    \multirow{2}{*}{French}  & ``je m'abonne''                & ``j'accepte''                      \\
                             & ``s’abonner''                  & ``accepter et continuer''          \\ \specialrule{.15em}{.01em}{.075em} 
    \end{tabular}
    \label{tab:keyphrases}
\end{table*}

\subsection{Exploratory Phase}


We conducted pilot crawls throughout the start of the study to create a set of HTML IDs and class names used for the detection of cookie paywalls as well as a corpus of syntagms\footnote{Sets of words with a sequential relationship to one another.} related to cookie paywalls. During these crawls, we also found different implementations of cookie paywalls where a user has to interact with the website (move the mouse or scroll) for a paywall to appear, or two-tiered cookie paywalls that present themselves as normal cookie banners until the user declines cookies. These findings were later used in the design of the automated detection approach of the full-scale crawler.

\subsection{Cookie Paywall Detection Approach}
To accommodate for the different types of implementations of cookie paywalls, the detection algorithm is run several times, with one action performed between each run. This program flow is illustrated in Figure~\ref{fig:flowchart_web_crawler}. 
A website gets flagged as not using a cookie paywall only if moving the mouse, scrolling, and trying to click a reject button did not result in any cookie paywall being detected. 

The detection algorithm parses the Document Object Model (DOM) representation of a website to search for web elements related to cookie paywalls by searching for HTML IDs and class names associated with cookie paywalls.  
A subset of the HTML attributes used by the four most prevalent CMPs in the dataset can be found in Table~\ref{tab:IDs&ClassNames}.
However, several websites make use of iframes and shadow DOM environments when implementing cookie paywalls, which cannot be accessed directly using Selenium. 
To work around this, the crawler searches for iframe tags and elements with the \textit{shadow\_root} property and switches its browsing context to the identified iframe or accesses the shadow DOM through the \textit{shadow\_root}.

\begin{table}[H]
    \small
    \centering
    \setcounter{table}{0}
    \caption{Example IDs and class names associated with cookie paywalls.}
    \begin{tabular}{@{}c|ll@{}}
    \specialrule{.15em}{.075em}{.01em} 
    Class/ID & \multicolumn{1}{c}{Attribute}        & \multicolumn{1}{c}{Associated CMP} \\ \specialrule{.08em}{.0em}{.0em} 
    \multirow{3}{*}{Class}  & cmp\_paywall                         & Traffective GmbH                    \\ \cline{2-3}
                            & didomi-popup-                        & \multirow{2}{*}{Didomi}             \\
                            & container                            &                                     \\ \hline
    \multirow{3}{*}{ID}     & \multirow{2}{*}{sp\_message\_iframe} & Sourcepoint                         \\
                            &                                      & Technologies                        \\ \cline{2-3}
                            & cmpwrapper                           & Consentmanager                      \\ \specialrule{.15em}{.01em}{.075em} 
    \end{tabular}
    \label{tab:IDs&ClassNames}
\end{table}
Elements found to relate to cookie paywalls are first searched for text confirming the presence of a cookie paywall using the corpus of syntagms.
Two syntagms are, in general, considered sufficient to identify a cookie paywall due to the two-choice nature of a cookie paywall, containing a payment option and a cookie option. If any of the syntagm combinations are found, the website is flagged as using a cookie paywall. Some representative combinations are outlined in Table~\ref{tab:keyphrases}. If no cookie paywall is found in the searched elements, the text in the entire DOM is searched using the same approach.

We measured the accuracy of cookie paywall detection on a subset of the dataset because of the infeasibility of a manual verification on the entire dataset. Instead, we performed a manual inspection on 10\% of randomly selected websites for each configuration of the crawler and found that the crawler produced false negatives for six cookie paywalls (three websites in two configurations). Thus, the crawler has an accuracy and recall of 99.7\% and precision of 100\%.

\subsection{TC String Collection and Analysis}
When a cookie paywall is found, the crawler retrieves the TC String, containing data on the purposes, features, and vendors a user has consented to,  
at two separate instances: before and after all cookies are accepted. The crawler's collection procedure is performed in three consecutive steps, where the success of one step ends the procedure:

\begin{itemize}
    \item[1)] The TCF API is leveraged via a JavaScript command.
    \item[2)] The cookie storage is searched for cookies containing the TC String.
    \item[3)] The browser's local storage is searched for entries containing the TC String. 
\end{itemize}

If found, the TC String is collected and decoded using the website \textit{iabtcf.com}~\cite{iabtcf_link} and the number of vendors to which data is conveyed based on both consent and legitimate interest are extracted. Additionally, the program extracts the purposes for data processing. 


\subsection{Double Paywall Detection Approach}

\begin{figure}[!ht]
    \small
    \centering
    \includegraphics[width=0.70\linewidth]{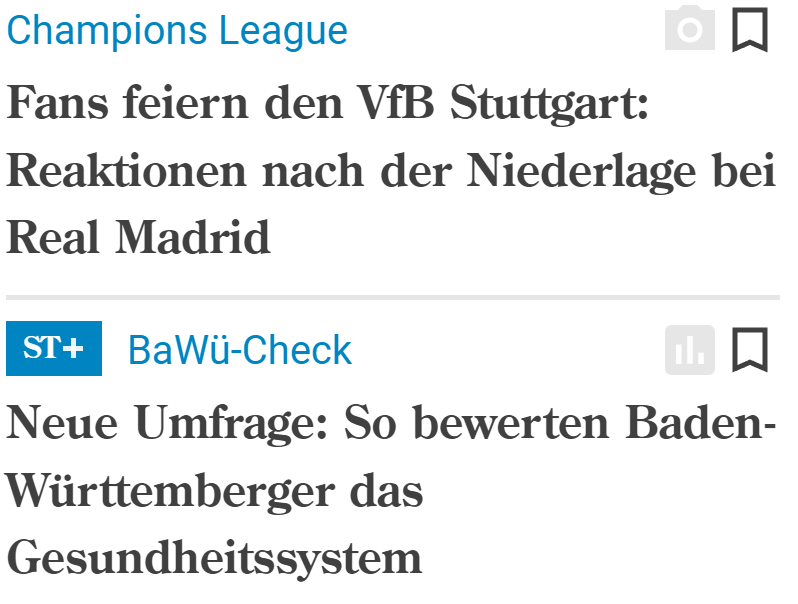}
    \caption{A Premium article for ``Stimme+'' subscribers under a free article on \url{https://www.stimme.de}.}
    \label{fig:dp_premium}
\end{figure}

After accepting cookies, the website can be interacted with and traversed in order to search for additional paywalls. These paywalls were only found on ``premium'' parts of websites, such as subscription-only articles, when exploring the dataset. 
For example, some websites indicate that an article is for subscription members only by including a ``premium icon'' with the class name \textit{premium-icon}. An example of this can be seen in Figure~\ref{fig:dp_premium}. 
If no elements with premium indicators are found, the crawler searches for any article, and as a last resort, any link that does not lead off the website.

\begin{figure*}[!b]
    \small
    \includegraphics[width=\linewidth]{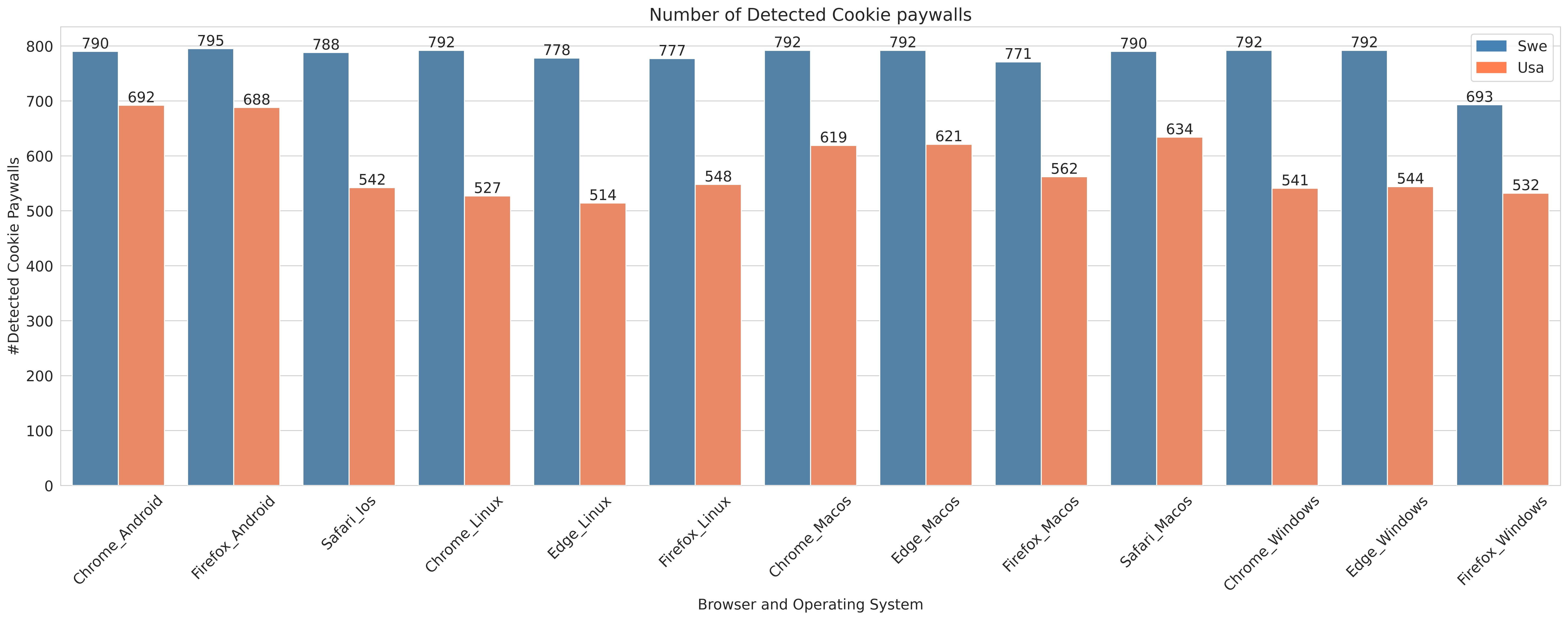}
    \caption{Number of cookie paywalls detected for each combination of the studied factors.}
    \label{fig:cp_detected_swe_usa}
\end{figure*}

The detection of additional paywalls is done heuristically in an ad hoc fashion through searching the webpage for visible elements that either have a class name or ID indicating the presence of a paywall, for example \textit{paywall} or \textit{paid-barrier}. 
If such an element is found, the crawler takes a screenshot of the element and saves the URL for manual inspection. If no potential paywalls are found on a webpage, the crawler returns to the previous page and searches for a new element.

This detection procedure is performed on up to three separate webpages on each website, where each webpage is examined only once. If a double paywall is detected or the maximum number of webpages are searched, the website has been successfully crawled, and the crawler proceeds to the next website in the dataset.

To verify the accuracy of the double paywall detection approach, the screenshots taken by the crawler were manually inspected. If a screenshot was inconclusive, the URL to the webpage that contained the additional paywall was examined. If the webpage did not present a paywall, the entire website was manually examined by traversing different webpages. The crawler identified 104 websites as presenting a double paywall, 93 of which had an actual double paywall, resulting in an accuracy of 88\%.

%% file: include/Results.tex
\section{\uppercase{Results}}
\label{sec:results}
This chapter presents the data collected from the websites in the dataset, see Section~\ref{sec:dataset}. The data was collected using the 26 possible configurations of the crawler, that is, combinations of browser, operating system and geographic location. 

\subsection{Prevalence of Cookie Paywalls}
\label{prevalence_cp}

Figure~\ref{fig:cp_detected_swe_usa} presents the number of cookie paywalls detected in each of the configurations. The highest number of cookie paywalls found in any configuration was 795, using Firefox on Android from the Swedish vantage point, and the lowest was 514 when using Edge on Linux from the USA. On average, 780 cookie paywalls were detected for each configuration in Sweden, and 581 for the configurations in the USA. Additionally, the number of detected cookie paywalls in Sweden was higher than that of the USA across all configurations of the crawler.

\begin{figure*}[ht]
    \small
    \centering
    \includegraphics[width=\linewidth]{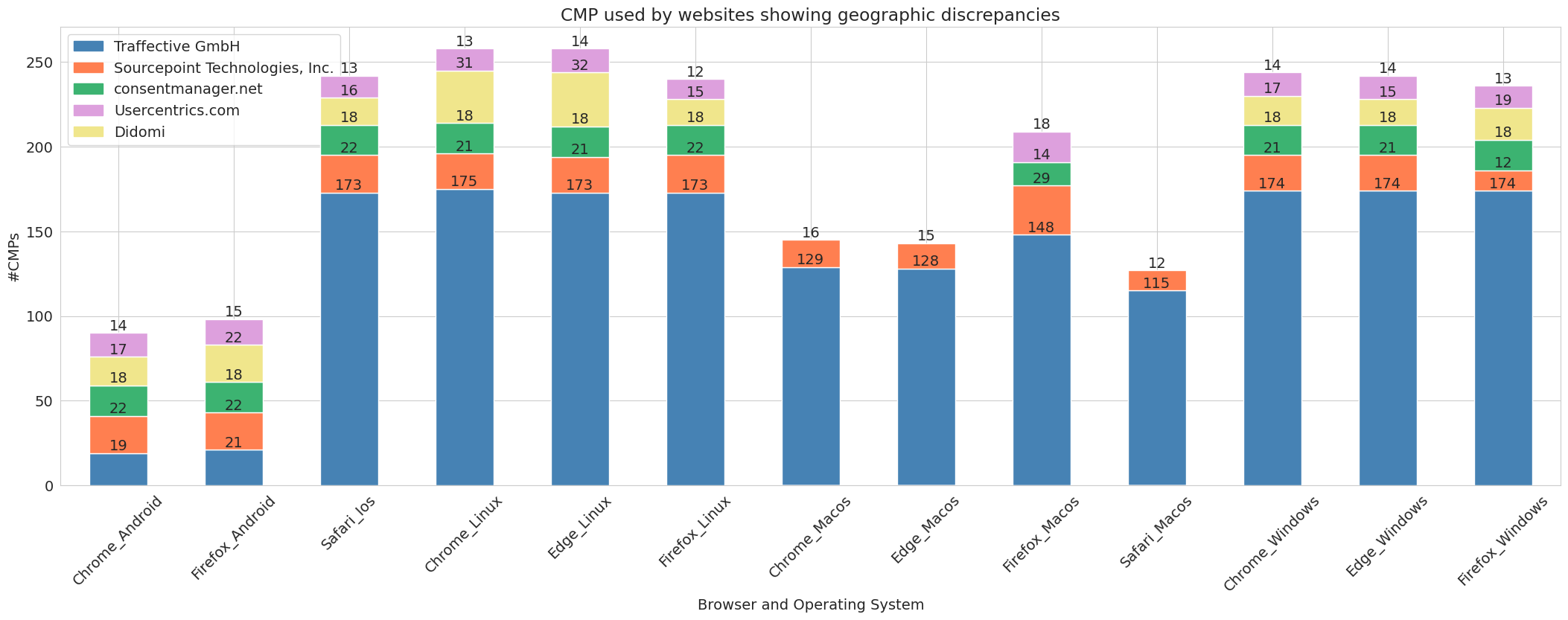}
    \caption{Distribution of CMPs used by websites that present a cookie paywall when accessed from the Swedish vantage point but not the USA. Only CMPs used on more than 10 websites are considered.}
    \label{fig:cmp_diff}
\end{figure*}

When considering the CMPs used by the websites, it was found that the CMP \textit{Traffective GmbH} was overrepresented among the websites presenting a cookie paywall in Sweden but not in the USA.
Figure~\ref{fig:cmp_diff} shows that this CMP was used by approximately 70\% of these websites for all browser and operating system combinations, except when using Android. When using an Android device, the CMP \textit{Traffective GmbH} was the second most prevalent CMP at 21\% after \textit{Sourcepoint Technologies Inc} at 23\%. 

\subsection{Actions Required to Display a Cookie Paywall}
\label{access_actions}
Cookie paywalls were detected by the crawler on 800 websites and produced complete data for 380 websites, that is, a cookie paywall was found in every configuration of these 380 websites.
To be properly able to compare the different combinations of the studied factors, the remainder of this section focuses on these 380 websites. 

The distribution of instances where a cookie paywall was presented a) immediately, b) after moving the mouse and c) after scrolling are presented in Table~\ref{tab:actions_dist_browser}, Table~\ref{tab:actions_dist_os} and Table~\ref{tab:actions_dist_location}. The tables show the distribution over each browser, operating system and geographic location, respectively.

\begin{table}[H]
    \centering
    \small
    \setcounter{table}{2}
    \caption{Distribution of required actions, aggregated over each browser.}
    \begin{tabular}{@{}l|ccc@{}}
    \specialrule{.15em}{.075em}{.01em} 
    Browser & Initial        & Mouse       & Scroll     \\ \specialrule{.08em}{.0em}{.0em}
    Chrome  & 3020 (99.34\%) & 19 (0.63\%) & 1 (0.03\%) \\
    Edge    & 2261 (99.17\%) & 18 (0.79\%) & 1 (0.04\%) \\
    Firefox & 3014 (99.14\%) & 22 (0.72\%) & 4 (0.13\%) \\
    Safari  & 1516 (99.74\%) & 4 (0.26\%)  & 0 (0\%)    \\ \specialrule{.15em}{.01em}{.075em} 
    \end{tabular}
    \label{tab:actions_dist_browser}
\end{table}

\begin{table}[H]
    \centering
    \small
    \caption{Distribution of required actions, aggregated over each operating system.}
    \begin{tabular}{@{}l|ccc@{}}
     \specialrule{.15em}{.075em}{.01em} 
    Operating & \multirow{2}{*}{Initial}        & \multirow{2}{*}{Mouse}       & \multirow{2}{*}{Scroll}     \\
    System    &                &             &            \\ \specialrule{.08em}{.0em}{.0em}
    Android   & 1513 (99.54\%) & 2 (0.13\%)  & 5 (0.33\%) \\
    iOS       & 758 (99.74\%)  & 2 (0.26\%)  & 0 (0\%)    \\
    Linux     & 2248 (98.60\%) & 31 (1.36\%) & 1 (0.04\%) \\
    macOS     & 3034 (99.80\%) & 6 (0.20\%)  & 0 (0\%)    \\
    Windows   & 2258 (99.04\%) & 22 (0.96\%) & 0 (0\%)    \\ \specialrule{.15em}{.01em}{.075em} 
    \end{tabular}
    \label{tab:actions_dist_os}
\end{table}

\begin{table}[H]
    \centering
    \small
    \caption{Distribution of required actions, aggregated over each geographic location.}
    \begin{tabular}{@{}l|ccc@{}}
    \specialrule{.15em}{.075em}{.01em} 
    Country & Initial        & Mouse       & Scroll     \\ \specialrule{.08em}{.0em}{.0em}
    Swe      & 4925 (99.70\%) & 15 (0.30\%) & 0 (0\%)    \\
    USA     & 4886 (98.91\%) & 48 (0.97\%) & 6 (0.12\%) \\ \specialrule{.15em}{.01em}{.075em} 
    \end{tabular}
    \label{tab:actions_dist_location}
\end{table}

\begin{table*}[!b]
    \centering
    \small
    \setcounter{table}{8}
    \caption{Statistics over the max difference in the number of vendors on a per-website basis.}
    \begin{tabular}{l|l|ccc}
        \specialrule{.15em}{.075em}{.01em} 
         Cookies & \multirow{2}{*}{Legal Basis} & Maximal & Mean & Standard \\
         Accepted & & Max Diff & Max Diff & Deviation \\
        \specialrule{.08em}{.0em}{.0em}
        \multirow{2}{*}{No}  & Consent & 0 & 0 & 0\\\cline{2-5}
          & Legitimate Interest & 3 & 0.12 & 0.43 \\\hline
        \multirow{2}{*}{Yes} & Consent & 26 & 1.21 & 2.53\\\cline{2-5}
         & Legitimate Interest & 6 & 0.30 & 0.88 \\
        \specialrule{.15em}{.01em}{.075em} 
    \end{tabular}
    \label{tab:tracking_vendors_diff}
\end{table*}

The vast majority of websites presented a cookie paywall immediately after entering the website, regardless of the browser, operating system and geographic location. In addition, scrolling was not necessary for any website when accessed from the Swedish vantage point, when browsing using the Safari browser or when using iOS, macOS, or Windows. Across all combinations, only 17 websites showed any kind of discrepancy in what action was required, and for each of these websites the cookie paywall appeared immediately in at least one configuration.

\subsection{Processing of User Data Conveyed Through the TC String}
\label{tracking}

The crawler produced complete data (a valid TC String was collected in every configuration) for 238 websites before accepting cookies and 335 websites after accepting cookies. The following results are derived from data collected on these 238 and 335 websites.

\subsubsection{Number of Vendors}
\label{sec:number_of_vendors}
The number of websites that showed any discrepancies in the number of registered vendors after accepting cookies were 124 and 23 for consent and legitimate interest, respectively. Before accepting cookies, no websites showed any discrepancies in the number of vendors using consent, and 16 websites showed discrepancies for vendors using legitimate interest.

Table~\ref{tab:vendors_consent_browser}, Table~\ref{tab:vendors_consent_os} and Table~\ref{tab:vendors_consent_location} show the average number of vendors that were registered on the basis of consent after cookies were accepted, as well as the percentual difference between the mean of \textit{each} configuration and the mean over \textit{all} configurations. The tables show the results aggregated over each browser, operating system and geographic location, respectively. 
No difference was shown on any website in the number of vendors using consent as legal basis.

\begin{table}[H]
    \centering
    \small
    \setcounter{table}{5}
    \caption{Statistics on the number of vendors using consent as a legal basis after accepting cookies, aggregated over each browser.}
    \begin{tabular}{@{}l|ccc@{}}
    \specialrule{.15em}{.075em}{.01em} 
    \multirow{2}{*}{Browser} & \multirow{2}{*}{Mean}    & Standard  & Diff     \\
            &         & Deviation & Mean     \\ \specialrule{.08em}{.0em}{.0em}
    Firefox & 280.082 & 274.633   & -0.054\% \\
    Chrome  & 280.225 & 274.582   & -0.003\% \\
    Edge    & 280.248 & 274.606   & 0.005\%  \\
    Safari  & 280.528 & 274.653   & 0.105\%  \\ \specialrule{.15em}{.01em}{.075em} 
    \end{tabular}
    \label{tab:vendors_consent_browser}
\end{table}
\begin{table}[H]
    \centering
    \small
    \caption{Statistics on the number of vendors using consent as a legal basis after accepting cookies, aggregated over each operating system.}
    \begin{tabular}{l|ccc}
        \specialrule{.15em}{.075em}{.01em} 
        Operating & \multirow{2}{*}{Mean} & Standard  & Diff \\
        System    &      & Deviation & Mean \\ 
        \specialrule{.08em}{.0em}{.0em}
        Linux & 280.063 & 274.589 & -0.061\% \\
        Windows & 280.080 & 274.612 & -0.055\% \\
        macOS & 280.429 & 274.639 & 0.070\% \\
        iOS & 280.634 & 274.729 & 0.143\% \\
        Android & 280.127 & 274.652 & -0.038\% \\
        \specialrule{.15em}{.01em}{.075em} 
    \end{tabular}
    \label{tab:vendors_consent_os}
\end{table}

\begin{table}[H]
    \centering
    \small
    \caption{Statistics on the number of vendors using consent as a legal basis after accepting cookies, aggregated over each geographic location.}
    \begin{tabular}{l|ccc}
        \specialrule{.15em}{.075em}{.01em} 
        \multirow{2}{*}{Country} & \multirow{2}{*}{Mean} & Standard  & Diff \\
                &      & Deviation & Mean \\
         \specialrule{.08em}{.0em}{.0em}
        Swe & 280.068 & 274.493 & -0.059\% \\
        USA & 280.399 & 274.673 & 0.059\% \\
        \specialrule{.15em}{.01em}{.075em} 
    \end{tabular}
    \label{tab:vendors_consent_location}
\end{table}

For both consent and legitimate interest, both before and after accepting cookies, the number of registered vendors varied greatly across different websites, with a smaller variation for the same website with different configurations.   
However, a slightly higher number of vendors were registered in the USA than in Sweden. The same was true for when using Edge and Safari compared to the other web browsers, and when using an Apple operating system (macOS or iOS) compared to the other studied operating systems.

When instead examining the difference between configurations on a per-website basis, it was found that if there was a difference between configurations, it tended to be small.
For example, the max difference in the number of vendors using consent as a basis for data processing was 26\footnote{Found on the website \textit{as.com} with 806 vendors on Firefox on Windows from the USA and 780 on Chrome on Linux from Sweden.}. 
However, the average max difference over all websites was 1.21, with a standard deviation of 2.53. The same information for consent and legitimate interest before and after cookies were accepted can be found in Table~\ref{tab:tracking_vendors_diff}. 

\subsubsection{Purposes for Data Processing}
\label{purposes_for_data_processing}

No website registered any purposes based on consent before cookies were accepted. After cookies were accepted, only one website registered different purposes based on consent depending on the configuration used on the crawler. On this website, fewer purposes were registered for all combinations involving Firefox. For legitimate interest, 16 and 37 websites differed between configurations before and after accepting cookies, respectively. In all cases where some discrepancy was found, the difference consisted of a varying number of purposes registered.

Table~\ref{tab:purposes_stats_browser}, Table~\ref{tab:purposes_stats_os} and Table~\ref{tab:purposes_stats_location} show the average number of allowed purposes for a vendor using legitimate interest, both before and after cookies were accepted. The tables show the results aggregated over each browser, operating system and geographic location, respectively. 

\begin{table}[H]
    \centering
    \small
    \setcounter{table}{9}
    \caption{Statistics on the number of purposes allowed for vendors using legitimate interest as a basis, aggregated over each browser.}
    \begin{tabular}{l|cc|cc}
       \specialrule{.15em}{.075em}{.01em} 
        \multirow{2}{*}{Browser} & \multicolumn{2}{c|}{Before}   & \multicolumn{2}{c}{After} \\
           & Mean & \multicolumn{1}{c|}{Diff Mean } & Mean & Diff Mean \\
        \specialrule{.08em}{.0em}{.0em}
        Firefox & 0.849 & -2.088\%  & 3.028 & -0.544\% \\
        Chrome & 0.870  & 0.336\% & 3.050  & 0.167\% \\
        Edge & 0.871 & 0.497\% & 3.046 & 0.040\% \\
        Safari & 0.891 & 2.759\% & 3.066 & 0.694\% \\
        \specialrule{.15em}{.01em}{.075em} 
    \end{tabular}
    \label{tab:purposes_stats_browser}
\end{table}

\begin{table}[H]
    \small
    \centering
    \caption{Statistics on the number of purposes allowed for vendors using legitimate interest as a basis, aggregated over each operating system.}
    \begin{tabular}{l|cc|cc}
       \specialrule{.15em}{.075em}{.01em} 
        Operating & \multicolumn{2}{c|}{Before}   & \multicolumn{2}{c}{After} \\
          System & Mean & \multicolumn{1}{c|}{Diff Mean} & Mean & Diff Mean\\
        \specialrule{.08em}{.0em}{.0em}
        Linux & 0.854 & -1.442\% & 3.043 & -0.058\% \\
        Windows & 0.854 & -1.442\%  & 3.043 & -0.058\% \\
        macOS & 0.882  & 1.790\% & 3.031 & -0.434\% \\
        iOS & 0.899 & 3.729\% & 3.101 & 1.870\% \\
        Android & 0.857 & -1.119\% & 3.048  & 0.106\% \\
        \specialrule{.15em}{.01em}{.075em} 
    \end{tabular}
    \label{tab:purposes_stats_os}
\end{table}

\begin{table}[H]
    \centering
    \small
    \caption{Statistics on the number of purposes allowed for vendors using legitimate interest as a basis, aggregated over each geographic location.}
    \begin{tabular}{l|cc|cc}
        \specialrule{.15em}{.075em}{.01em} 
        \multirow{2}{*}{Country} & \multicolumn{2}{c|}{Before}   & \multicolumn{2}{c}{After} \\
          & Mean & \multicolumn{1}{c|}{Diff Mean} & Mean & Diff Mean \\
         \specialrule{.08em}{.0em}{.0em}
        Swe & 0.860 & -0.746\% & 2.945 & -3.258\% \\
        USA   &  0.873 & 0.746\% & 3.144 & 3.258\% \\ 
        \specialrule{.15em}{.01em}{.075em} 
    \end{tabular}
    \label{tab:purposes_stats_location}
\end{table}

\begin{figure*}[!ht]
\begin{minipage}{.5\textwidth}
 \small
 \hspace*{-0.4cm}
  \includegraphics[width=\linewidth]{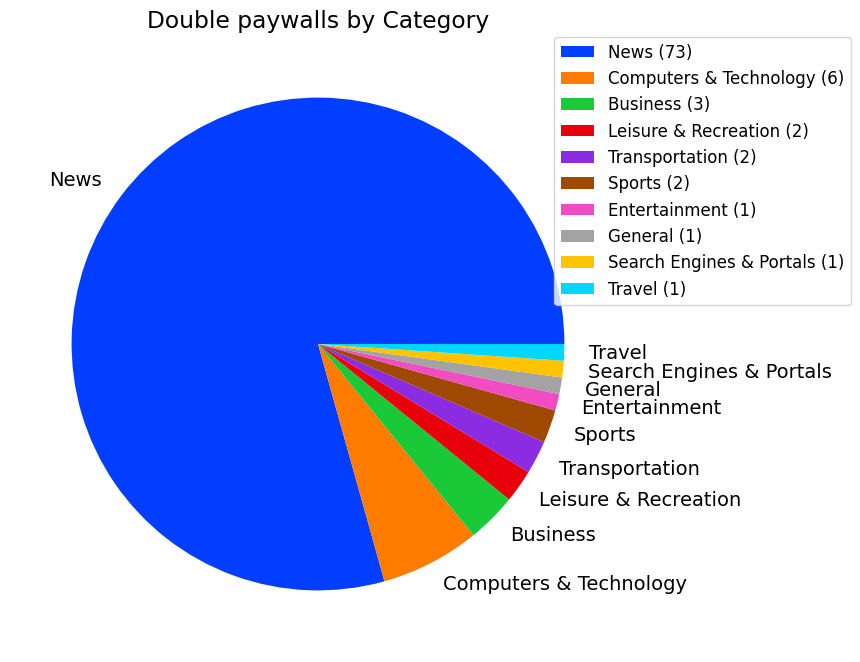}
  \caption{Double paywalls by Category.}
  \label{fig:dp_category}
\end{minipage}%
\begin{minipage}{.5\textwidth}
  \small
  \hspace*{0.6cm}
  \includegraphics[width=0.86\linewidth]{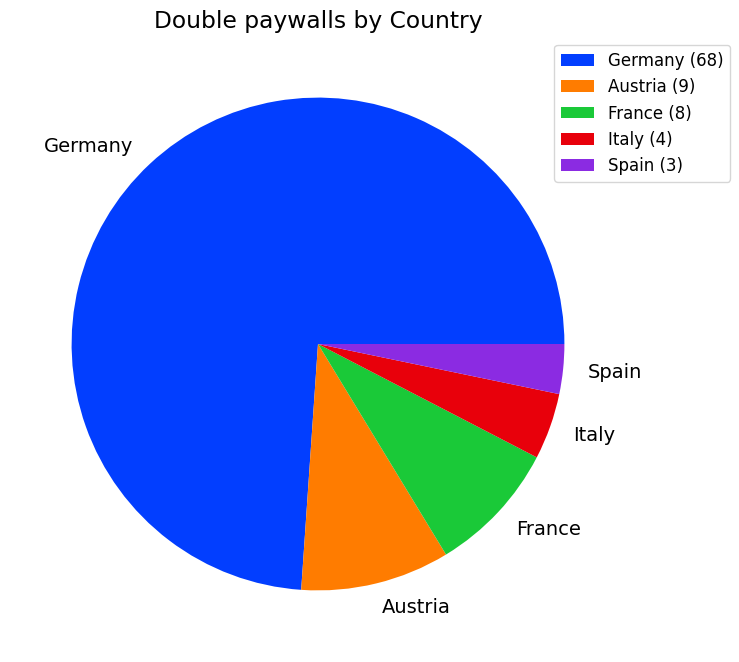}
  \caption{Double paywalls by Country.}
  \label{fig:dp_country}
\end{minipage}

\end{figure*}

When accessing websites from Sweden, the average number of purposes registered was lower than when accessing from the USA. Furthermore, it was found that 20 websites used legitimate interest as a basis to select or create a profile for personalized ads or content in the USA, whereas no websites did this in the Swedish vantage point. When including the websites where complete data could not be obtained, this number increases to 39. All of these websites used \textit{consentmanager.net} as their CMP. 

Devices using Firefox did, on average, register fewer purposes for vendors using legitimate interest as their basis. Additionally, devices using iOS or macOS registered more purposes before cookies were accepted. After cookies were accepted, mobile devices (Android and iOS) registered more purposes than desktop devices.

\subsection{Prevalence of Double Paywalls}
\label{double_paywalls}
We found 93 websites\footnote{\url{https://github.com/Trooja/cookie-paywall-discrepancies/blob/main/double-paywalls.csv}}
that presented a double paywall.
Out of these, 73 websites were classified as news websites according to Cyren’s URL Category Checker~\cite{cyren_website_2024}, followed by the categories \textit{Computers \& Technology} and \textit{Business}. A more detailed categorization can be found in Figure~\ref{fig:dp_category}, in which each category is presented with the number of websites classified as said category.  

Out of the detected double paywalls, the majority of websites were hosted in Germany, as can be seen in Figure~\ref{fig:dp_country}. Additionally, all double paywalls were hosted in one of five European countries that are members of the EU, with Austria and France having the second and third highest number of double paywalls.

%% file: include/Discussion.tex
\section{\uppercase{Discussion}}

This section discusses the results presented in Section~\ref{sec:results} followed by limitations of the study and suggestions for future work. 

\subsection{Prevalence of Cookie Paywalls}
\label{dis:prevelence}

\textbf{Out of the studied factors, the geographic location has the largest impact on whether a website presents a cookie paywall or not. }
For all combinations of browser and operating system, fewer cookie paywalls were detected when accessing websites from the USA than when accessing them from Sweden.
It was found that, in all but two combinations, the majority of websites showing this type of behavior used the CMP \textit{Traffective GmbH}. 
Thus, it is possible that this CMP examines the location of the user and almost systematically chooses to not present a cookie paywall if the user is located outside the EU. 
However, six websites using this CMP did not exhibit this behavior in any of the combinations, indicating the presence of other, possibly website-related, factors.
Additionally, the CMP was only used by 20\% of the websites displaying this behavior on the two combinations using Android, which would suggest that the choice of browser and operating system is also of some importance. 

\textbf{The type of device (mobile or desktop) accessing a website does not appear to be directly correlated to the prevalence of cookie paywalls.}
There were larger differences in the number of detected cookie paywalls between the two mobile devices than the difference between the iOS device and any of the desktop operating systems. However, when studying differences between all \textit{operating systems}, one can see that using Android, especially in the USA, resulted in more cookie paywalls being encountered. 
Thus, the role of the operating system seems to be of greater importance than the type of device used. Furthermore, the operating system seems to play a bigger role when accessing websites from the USA than when accessing them from Sweden. 

\textbf{In general, the choice of web browser does not seem to impact the prevalence of cookie paywalls on its own.} In the majority of cases, fewer cookie paywalls were encountered when using Firefox compared to the other browsers, but the average number of cookie paywalls did not differ by more than 25 websites. \textbf{However, one configuration of the crawler produced an outlier, namely when using Firefox on Windows from Sweden.} This combination presented 78 fewer cookie paywalls than the second lowest encounter rate in Sweden, and only one more cookie paywall than the highest encounter rate in the USA. The reason for this outlier is unknown but may partly be a result of the CMP \textit{Sourcepoint Technologies}, as this CMP was used by approximately 67\% of the websites with no cookie paywall in this configuration.

\textit{\textbf{To summarize:}
The prevalence of cookie paywalls was most affected by the geographic location used to access them, with more cookie paywalls being displayed from the Swedish vantage point than the USA. The browser had little effect on the number of cookie paywalls encountered, one combination of all three factors produced an outlier. The device type did not seem to directly correlate to the prevalence of cookie paywalls, but the operating system of the device did.
}

\subsection{Actions Required to Display a Cookie Paywall}
\label{dis:actions}

\textbf{The vast majority of cookie paywalls appeared immediately, and few websites required that the crawler moved the mouse before the cookie paywall was displayed.} 
Only 17 websites, approximately 4\% of the websites with complete data, displayed differences in the required action, or lack thereof, when accessed using different configurations of the crawler. This small number of websites showing any kind of discrepancy makes it infeasible to determine if these differences depend on the studied factors or inconsistencies in the behavior of the crawler, network latency or some other factor. 
Consequently, it is plausible that these discrepancies are website specific rather than a result of the different configurations used. 

\textit{\textbf{To summarize:}
98\% of displayed cookie paywalls appeared immediately, and only 17 websites required any type of action for the cookie paywall to appear. 
}

\subsection{Processing of User Data Conveyed Through the TC String}
\label{dis:tracking}
The analysis of the number of vendors and purposes for the different legal bases was solely conducted on websites that provided this information for every combination. This choice enabled the creation of comparable results between configurations, but may also have led to potential patterns existing in the entire dataset being missed.

\textbf{Discrepancies in the number of vendors between different combinations of browser, operating system and geographic location were found, but the differences were too small to make any concrete connection to a specific factor.}
On a per-website basis, some differences were discovered between the combinations of the studied factors, but these differences were, in general, small. Furthermore, when comparing the averages aggregated over each factor, one can see that no single factor resulted in a significant deviation from the total mean, with the largest deviation being 0.34\%. These deviations were a result of a small subset of websites that on average differed by one vendor.

\textbf{The number of registered purposes varied between different combinations of the studied factors, but too few websites showed any discrepancies to distinguish a pattern.}
Only one website displayed discrepancies between different configurations of the crawler when consent was used as a legal basis.
When instead examining legitimate interest as a legal basis, we found 16 and 37 websites that showed some kind of discrepancy in what purposes were registered before and after accepting cookies, respectively. In all of these cases, the difference was constituted by the extension of the list of allowed purposes. Out of the studied factors, the geographic location seems to make the largest impact, registering more purposes in the USA than in Sweden, but because of the low number of websites presenting any type of discrepancy this pattern does not seem to be part of a larger trend.

\textbf{When accessed from the USA, 39 websites used legitimate interest as a basis to select or create a profile for personalized ads and content, but no websites did so when accessed from the Swedish vantage point.} The use of this legal basis for these purposes is forbidden in the current version of the TCF framework (TCF v2.2) after a decision by the Belgian Data Protection Authority in 2022~\cite{iab_europe_faq_2023}. This indicates that websites using \textit{consentmanager.net} as its CMP may, based on the geographic location, disregard the rules of the IAB Europe TCF and change how the collected user data is processed. 

\textit{\textbf{To summarize:} Discrepancies were found in the number of vendors user data was shared with, and the purposes for which these vendors processed data, but the differences were small and were only found on a subset of the websites. On these websites, a slight tendency for more purposes was displayed when accessed from the USA.}


\subsection{Prevalence of Double Paywalls}
\label{dis:double_paywalls}

\textbf{Approximately 11.6\% of websites used a double paywall.}
The majority of these websites (78\%) were categorized as news websites, including several large news sites such as \textit{zeit.de}, \textit{lemonde.fr} and \textit{abc.es}. This is significantly higher than the proportion of news sites among websites using cookie paywalls found in previous research~\cite{morel_legitimate_2023}. A possible explanation could be that this practice lends itself well to news websites, as it may allow a website to entice users to subscribe to the newspaper through providing a subset of its content, whilst still being able to monetize this audience through targeted advertising. However, the predominance of news websites may instead partially be an effect of a bias in the methodology. The detection of double paywalls was automated, using a limited number of indicators that an additional paywall was used. Thus, extending the set of indicators could provide a less biased search and potentially a wider variety of website categories.

The majority of double paywalls were found on websites based in Germany (73\%), followed by Austria (9.7\%) and France (8.6\%). The high prevalence of German websites can be explained by the majority (79\%) of cookie paywalls being found on German websites (one condition for a double paywall is that there is a cookie paywall). However, the distribution of double paywalls was not proportionate to the distribution of cookie paywalls. For example, a relatively small proportion of cookie paywalls in Germany were double paywalls (11\%) compared to Austria (35\%) and France (21\%). A possible explanation for this may be the disproportionate number of German websites in the dataset compared to any other country. 

\textit{\textbf{To summarize:}
The use of double paywalls is fairly widespread, with 93 websites using one. The majority of these were classified as news websites and found in Germany.
}

\subsection{Limitations}
\label{limitations}
Our study provides insight into what discrepancies can be found in cookie paywalls between different browsers, operating systems and geographic locations. However, it is important to consider certain limitations:
first, an automated approach was used to collect data for the analysis of the study. A subset of this data was manually examined to provide a degree of confidence for the conclusions, but this manual verification might not guarantee complete accuracy for all the collected data.
Second, websites have previously been proven to be able to detect Selenium-driven web browsers and consequently alter their behavior~\cite{cassel_omnicrawl_2022}. It has also been shown that websites may deliver different content to detected web bots than to normal users~\cite{jonker_fingerprint_2019}. Because of this, the crawler may not fully represent the regular website behavior of an actual user. 
Finally, the data for this study was collected over several weeks. Thus, there exists a possibility that some websites would have changed their behavior during the time of the data collection.


\subsection{Future Work}
\label{future_work}


Future studies may want to use a larger set of geographic locations and a wider range of browsers on common operating systems. A larger set of vantage points, both located inside and outside the EU, 
would give deeper insights into whether the trends found in this study apply solely to individual countries or extend to larger regions. Studying additional browsers, especially browsers such as Chrome and Firefox on iOS, can widen the basis for comparison between mobile and desktop devices. 

A further continuation of this work would be to explore not only the presence of cookie paywalls, but also what, if anything, replaces a cookie paywall on a website if it appears in some configurations but not in others. Additionally, a comparison of the tracking conducted by such websites could provide an understanding of how the collection of consent affects how user data is collected and used. 

%% file: include/Conclusion.tex
\section{\uppercase{Conclusion}}
\label{sec:conclusion}

We presented in this paper the first study on the effects of the web browser, device type and geographic location on the presence and behavior of cookie paywalls, and their handling of users' data. 
We built an automated crawler to collect data from 804 websites with confirmed cookie paywalls. Using this data, we showed that all factors affected cookie paywalls to some degree -- the location being the most impactful --, and that changing the combination of the factors predominantly affected the presence of the cookie paywall. Finally, we produced the first dataset of a new type of paywall coined \textit{double paywall}.
These last results were presented during a panel discussion organized by the EDPS in July 2024. \footnote{\url{https://www.edps.europa.eu/data-protection/our-work/publications/events/2024-07-11-consent-or-pay-how-can-single-market-and-fundamental-rights-work-together}} 
